\documentclass[preprint]{aastex}

\shortauthors{Wright}

\shorttitle{$w$ \& $w^\prime$}

\newcommand{\etal}         {{\it et al.}}
\newcommand{\vs}           {{\it vs.}}
\newcommand{\Geff}         {\Gamma_{\mbox{{\small eff}}}}
\newcommand{\be}           {\begin{equation}}
\newcommand{\ee}           {\end{equation}}
\newcommand{\bea}          {\begin{eqnarray}}
\newcommand{\eea}          {\end{eqnarray}}

\slugcomment{Draft: April 20, 2007}

\begin{document}

\title{Constraints on Dark Energy from Supernovae, $\gamma$-ray
bursts, Acoustic Oscillations, Nucleosynthesis and Large Scale
Structure and the Hubble constant}

\author{
E.\ L.\ Wright\altaffilmark{1},
}

\altaffiltext{1}{UCLA Astronomy, PO Box 951547, Los Angeles CA 90095-1547, USA}

\email{wright@astro.ucla.edu}

\begin{abstract}
The luminosity distance \vs\ redshift law is now measured using
supernovae and $\gamma$-ray bursts, and the angular size distance
is measured at the surface of last scattering by the CMB and at $z
= 0.35$ by baryon acoustic oscillations.  In this paper this data
is fit to models for the equation of state with $w = -1$, 
$w =\;$constant, and $w(z) = w_0+w_a(1-a)$.  The last model is poorly
constrained by the distance data, leading to unphysical solutions
where the dark energy dominates at early times unless the large
scale structure and acoustic scale constraints are modified to allow
for early time dark energy effects.  A flat $\Lambda$CDM model
is consistent with all the data.

\end{abstract}


%\newpage

\section{INTRODUCTION}\label{intro}

Wood-Vasey \etal\ (2007) recently published supernova data from the
ESSENCE project, while Riess \etal\ (2007) have published a large sample
of supernovae from the SNLS project (Astier \etal\ 2006),
the HST, and Hi-z Supernova Team.
Schaefer (2006) has published a sample of $\gamma$-ray burst
distances.  While GRBs give much less accurate distances than 
supernovae, they extend to much higher redshifts and the GRB
data helps to distinguish between non-flat geometries and
equations of state with $w \neq -1$ which both affect the
distance-redshift law at ${\cal O}(z^3)$ for low redshift.

The analysis of Wood-Vasey \etal\ (2007) plotted contours
of $w$ and $w_a$, based only on a subset of the supernova
data used here and a prior on $\Omega_M$.  These contours
extended into the region where the dark energy dominated the
density at the surface of last scattering or during nucleosynthesis.
Li \etal\ (2006) added GRBs and large scale structure data
to the SNe data but still show contours extending into the
early dark energy domination region.
Both Barger \etal\ (2006) and Riess \etal\ (2007)
used $w(z)$ laws in which the deviation of $w$ from $-1$
was terminated for $z > 1.8$, beyond the redshift of the most
distant supernova in the sample, but this is an arbitrary limit
which would have to modified to allow for GRBs.
Alam, Sahni \& Starobinsky (2006) analyzed the SNe data using
a quadratic polynomial in $(1+z)$ as the form for $\rho_{DE}$.
This form for $\rho_{DE}$ will never be dominant at early
times since the matter density varies like $(1+z)^3$.
Davis \etal\ (2007) have analyzed a subset of the combined 
supernovae, and have used an approximation to the CMB
acoustic peak constraint that fails when dark energy dominates
at high $z$.
The part of parameter space where dark energy dominates at
high $z$ obviously should be excluded, 
and I show in this paper that appropriate modifications to the
standard formulae for the acoustic scale, the $\Gamma$ parameter,
and Big Bang nucleosynthesis (BBNS) to allow
for the possible importance dark energy at $z \approx 10^9$
or $z \approx 10^3$ will lead to this exclusion automatically. 
The BBNS, acoustic scale and $\Gamma$ limits are given
in general forms involving $\rho_{DE}(z)$, which can be
used for any form of the equation of state.  But the
$w = w_\circ + w_a(1-a)$ form adopted by the
Dark Energy Task Force report (Albrecht \etal\ 2006) is used for
all the specific plots in this paper.  This form can be
used for all redshifts without introducing a redshift cutoff
which would be a new and otherwise unnecessary parameter.

\begin{table}[t]
\caption{Mean distance modulus relative to a Milne model for the 358
supernovae in Riess \etal\ (2007) Gold+Silver and Wood-Vasey
\etal\ (2007).\label{tab:SNall}}
\begin{center}
\begin{tabular}{rrrrr}
\tableline
\multicolumn{1}{c}{$\langle z \rangle$} & 
\multicolumn{1}{c}{$ \langle \Delta\mu \rangle $} & $z_{min}$ & 
$z_{max}$ & $N$ \\
\tableline
0.0159 & $ -0.052 \pm  0.118$ & 0.007 & 0.024 &  37 \\
0.0376 & $  0.004 \pm  0.057$ & 0.024 & 0.058 &  37 \\
0.0947 & $  0.103 \pm  0.075$ & 0.061 & 0.160 &  12 \\
0.2207 & $  0.110 \pm  0.062$ & 0.172 & 0.268 &  14 \\
0.3299 & $  0.096 \pm  0.037$ & 0.274 & 0.371 &  36 \\
0.4222 & $  0.170 \pm  0.037$ & 0.374 & 0.455 &  37 \\
0.4841 & $  0.245 \pm  0.037$ & 0.459 & 0.511 &  37 \\
0.5530 & $  0.169 \pm  0.034$ & 0.514 & 0.610 &  37 \\
0.6550 & $  0.100 \pm  0.036$ & 0.612 & 0.710 &  31 \\
0.7747 & $  0.054 \pm  0.053$ & 0.719 & 0.818 &  21 \\
0.8590 & $  0.055 \pm  0.064$ & 0.822 & 0.910 &  20 \\
0.9661 & $  0.047 \pm  0.068$ & 0.927 & 1.020 &  21 \\
1.1140 & $  0.017 \pm  0.118$ & 1.056 & 1.140 &   4 \\
1.2228 & $ -0.087 \pm  0.127$ & 1.190 & 1.265 &   5 \\
1.3353 & $ -0.151 \pm  0.100$ & 1.300 & 1.390 &   6 \\
1.4000 & $  0.037 \pm  0.810$ & 1.400 & 1.400 &   1 \\
1.5510 & $ -0.490 \pm  0.320$ & 1.551 & 1.551 &   1 \\
1.7550 & $ -0.599 \pm  0.350$ & 1.755 & 1.755 &   1 \\
\tableline
\end{tabular}
\end{center}
\end{table}

\section{OBSERVATIONS \label{sec:obs}}

\subsection{Supernovae \label{sec:SNe}}

\begin{table}[t]
\caption{Mean distance modulus relative to a Milne model for 272
supernovae in Riess \etal\ (2007) Gold and Wood-Vasey
\etal\ (2007).\label{tab:SNnoAg}}
\begin{center}
\begin{tabular}{rrrrr}
\tableline
\multicolumn{1}{c}{$\langle z \rangle$} & 
\multicolumn{1}{c}{$ \langle \Delta\mu \rangle $} & $z_{min}$ & 
$z_{max}$ & $N$ \\
\tableline
0.0169 & $ -0.035 \pm  0.125$ & 0.010 & 0.025 &  29 \\
0.0361 & $  0.024 \pm  0.067$ & 0.025 & 0.053 &  29 \\
0.0776 & $  0.074 \pm  0.080$ & 0.056 & 0.124 &  10 \\
0.2032 & $  0.123 \pm  0.073$ & 0.159 & 0.249 &  10 \\
0.3196 & $  0.105 \pm  0.042$ & 0.263 & 0.363 &  27 \\
0.4149 & $  0.144 \pm  0.039$ & 0.368 & 0.450 &  29 \\
0.4808 & $  0.203 \pm  0.040$ & 0.455 & 0.508 &  29 \\
0.5514 & $  0.139 \pm  0.039$ & 0.510 & 0.604 &  29 \\
0.6475 & $  0.119 \pm  0.038$ & 0.610 & 0.707 &  25 \\
0.7888 & $  0.058 \pm  0.059$ & 0.730 & 0.830 &  18 \\
0.8666 & $  0.007 \pm  0.081$ & 0.832 & 0.905 &  10 \\
0.9696 & $  0.035 \pm  0.076$ & 0.935 & 1.020 &  14 \\
1.1140 & $  0.017 \pm  0.118$ & 1.056 & 1.140 &   4 \\
1.2197 & $  0.081 \pm  0.143$ & 1.199 & 1.230 &   3 \\
1.3410 & $ -0.163 \pm  0.105$ & 1.300 & 1.390 &   5 \\
1.7550 & $ -0.599 \pm  0.350$ & 1.755 & 1.755 &   1 \\
\tableline
\end{tabular}
\end{center}

\end{table}

\begin{table}[t]
\caption{Mean distance modulus relative to a Milne model for 69
GRBs in Schaefer (2007).\label{tab:GRB}}
\begin{center}
\begin{tabular}{rrrrr}
\tableline
\multicolumn{1}{c}{$\langle z \rangle$} & 
\multicolumn{1}{c}{$ \langle \Delta\mu \rangle $} & $z_{min}$ & 
$z_{max}$ & $N$ \\
\tableline
0.2100 & $  0.453 \pm  0.362$ & 0.170 & 0.250 &   2 \\
0.4967 & $  0.458 \pm  0.289$ & 0.430 & 0.610 &   3 \\
0.7350 & $  0.479 \pm  0.217$ & 0.650 & 0.830 &   8 \\
0.9187 & $ -0.090 \pm  0.171$ & 0.840 & 1.020 &   8 \\
1.2083 & $  0.173 \pm  0.263$ & 1.060 & 1.310 &   6 \\
1.5275 & $ -0.202 \pm  0.175$ & 1.440 & 1.620 &   8 \\
2.0217 & $ -0.141 \pm  0.257$ & 1.710 & 2.200 &   6 \\
2.4825 & $ -0.117 \pm  0.211$ & 2.300 & 2.680 &   8 \\
3.1463 & $ -0.785 \pm  0.215$ & 2.820 & 3.370 &   8 \\
3.8243 & $ -0.573 \pm  0.247$ & 3.420 & 4.270 &   7 \\
4.6033 & $ -0.946 \pm  0.412$ & 4.410 & 4.900 &   3 \\
6.4450 & $ -1.100 \pm  0.463$ & 6.290 & 6.600 &   2 \\
\tableline
\end{tabular}
\end{center}
\end{table}

\begin{figure}[t]
\plotone{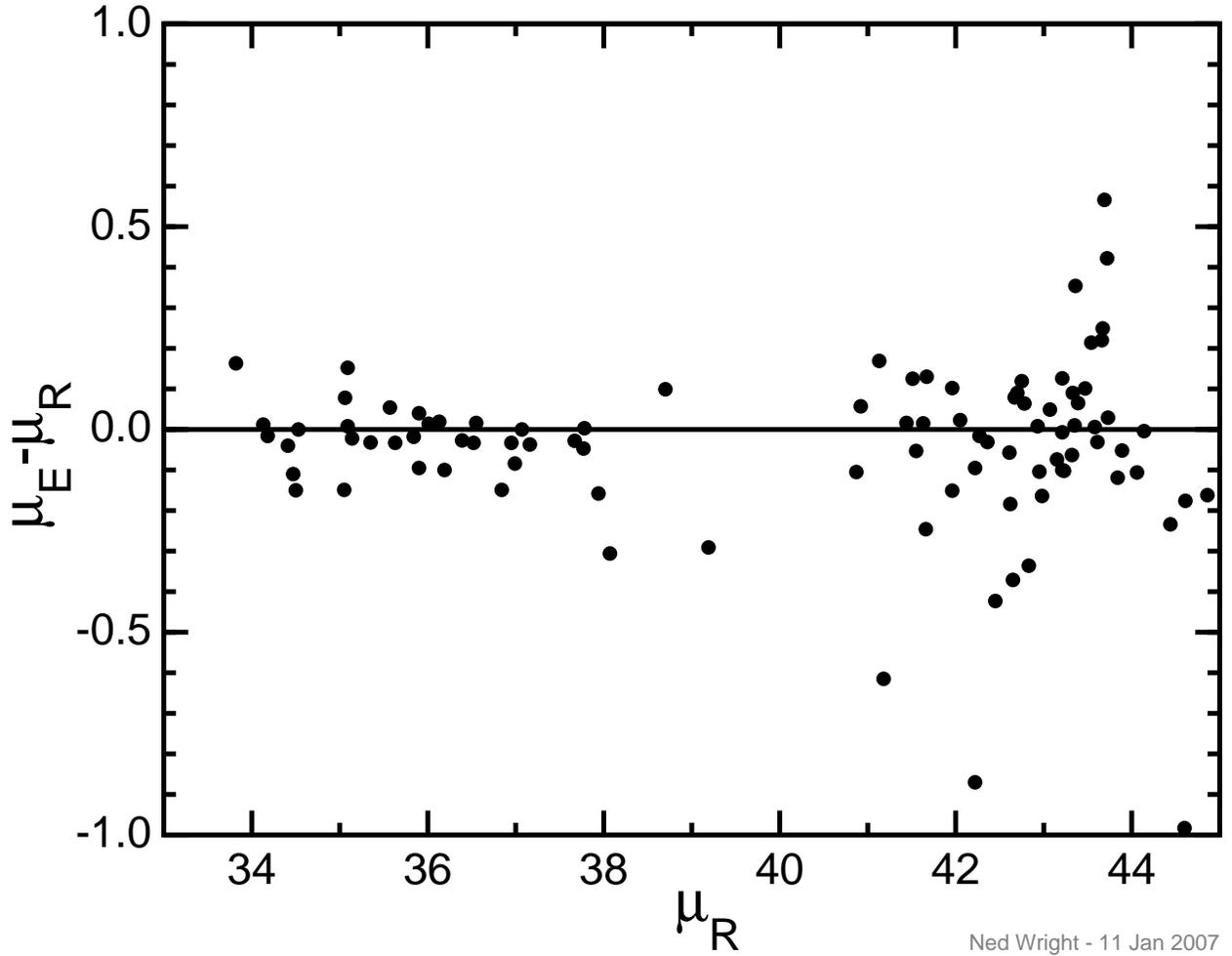}
\caption{Difference in distance moduli for objects common
to Wood-Vasey \etal\ (2007) ($\mu_E$) and Riess \etal\ (2007) ($\mu_R$).
The RMS scatter is 0.2 mag.\label{fig:ESSENCEtab9-Riess}}
\end{figure}

The distance modulus \vs\ redshift data from Riess \etal\ (2007) were
taken from the Web site provided by Riess.  The distance moduli and
redshifts for the ESSENCE supernovae were extracted from Table 9 in the
Latex file for astro-ph/0701041 (Wood-Vasey \etal\ 2007).
Typically different groups analyze
supernovae with different assumptions about the Hubble constant or
equivalently the absolute magnitude ${\cal M}$ of a canonical SN Ia
with a nominal decay rate.  In order to combine the new supernovae from
ESSENCE with the Riess \etal\ sample, it was necessary to check the
relative normalization of the two data sets using the 93 objects they
have in common.  Figure \ref{fig:ESSENCEtab9-Riess} shows the
comparison.  The scatter in the differential distance moduli is $0.2$
mag 1$\sigma$, which seems unusually high, and the median difference in
$\mu$ is 0.022 mag which is consistent with the standard deviation of
the mean given the scatter.  Objects which are not in the Riess
\etal\ sample but which had successful fits with $\chi^2$ per degree of
freedom $< 7$ were added to the Riess \etal\ sample, with the 0.022 mag
added to $\mu$, and an intrinsic scatter of $0.10$ mag added in
quadrature to $\sigma_\mu$.   This gives a total sample of 358 SNe.  I
have binned the SNe into bins containing $< 2+N_{tot}/10$ objects and
widths less than 0.1 in redshift.  An empty Universe model (the Milne
model) was first subtracted from the $\mu$ values.  The binned values
are listed in Table \ref{tab:SNall}.   A Hubble constant of 63.8
km/sec/Mpc was used when computing the Milne model, but this value has
no effect on the parameter limits computed in this paper.  Its only
effect is to add a constant to the $ \langle \Delta\mu \rangle $ values
in the Tables.  The mean differential distance modulus in each bin is
found by minimizing a modified
\be
\chi^2 = \sum_i f((<\Delta\mu>-[\mu_i-\mu_{Milne}])/\sigma_i)
\label{eq:modchi}
\ee
where $f(x) = x^2$ for $|x| < 2$, or  $4|x|-4$ otherwise.
The modification deweights extreme outliers.
Riess \etal\ (2007) recommend dropping SNe with redshifts
less than 0.023 to avoid a possible ``Hubble bubble'' seen by
Jha \etal (2007), but I have instead used a large velocity
error of $\sigma_v = 1500$ km/sec which gives an extra
$\sigma_\mu$ of $(5/\ln 10)(\sigma_v/cz)$ which is added
in quadrature with the tabulated $\sigma_\mu$.

Table \ref{tab:SNnoAg} was constructed the same way
but omitting the ``Silver'' objects in Riess \etal\ (2007).

\begin{figure}[t]
\plotone{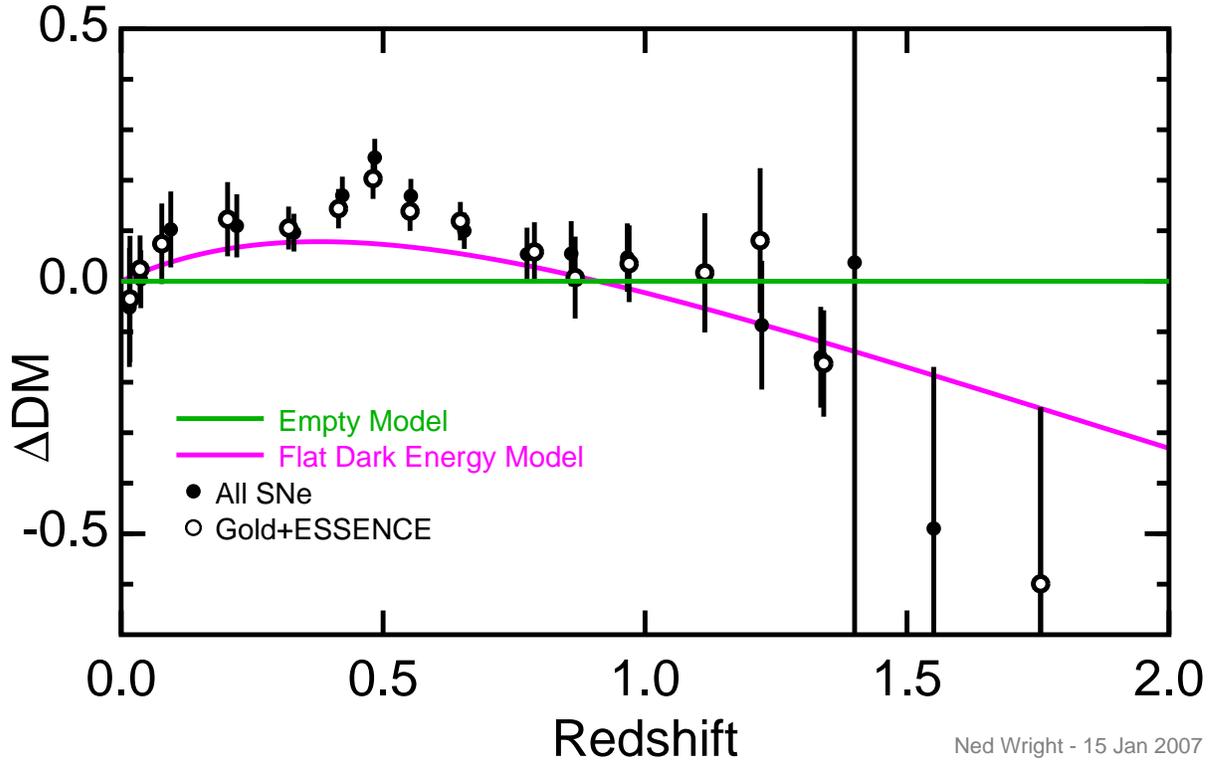}
\caption{Binned supernova data \vs\ redshift compared to
a flat $\Lambda$CDM model with $\Omega_M = 0.369$.
The filled circles are binned points from the full dataset,
while the open circles have omitted the ``Silver'' subset. 
\label{fig:dDM-vs-z-15Jan07}}
\end{figure}

The binned values from Tables \ref{tab:SNall} and  \ref{tab:SNnoAg} are
shown in Figure \ref{fig:dDM-vs-z-15Jan07} along with the flat
$\Lambda$CDM model that best fits the Hubble diagram data alone.  This
model has $\Omega_M = 0.369$.  There is an excursion around $z = 0.5$
that can be seen clearly in the binned supernova data.  A simple 3 parameter
fit to this bump gives a $\Delta\chi^2$ of 15 in the total sample, but only
6 if the ``Silver'' SNe are excluded.

\subsection{GRBs}

\begin{figure}[t]
\plotone{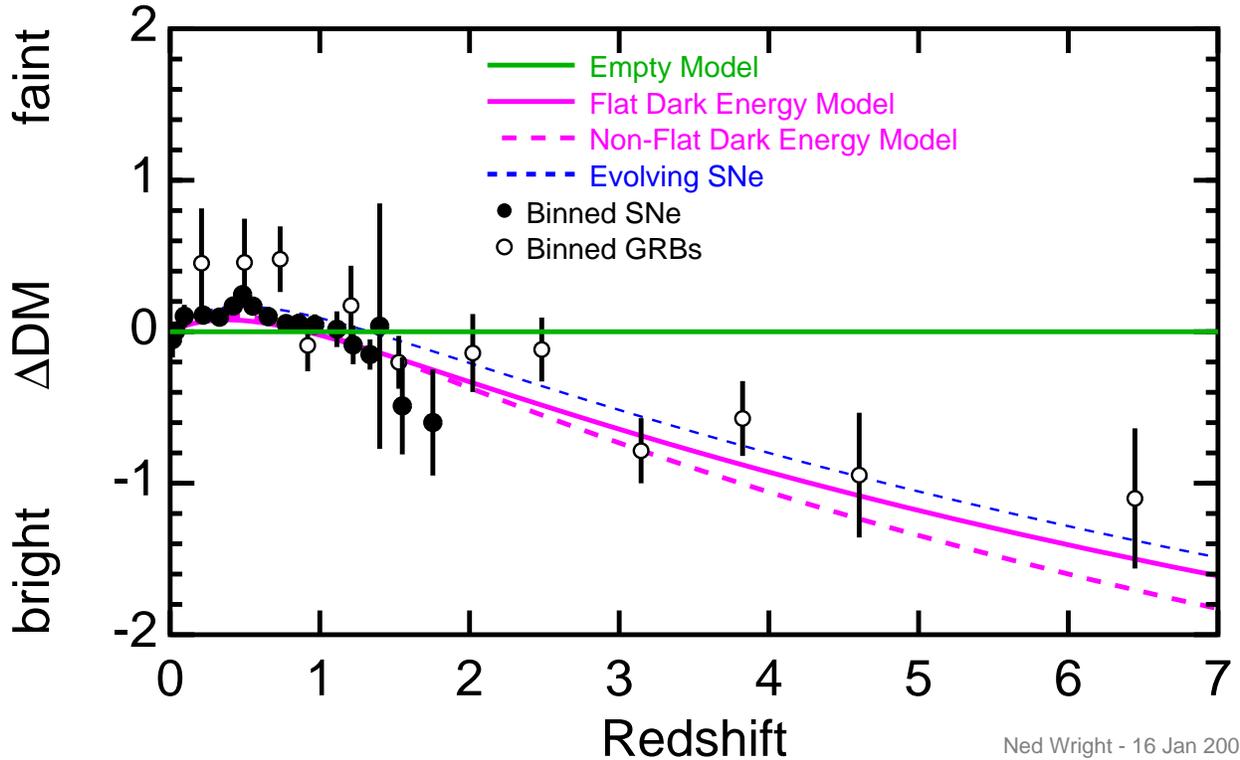}
\caption{Binned supernova and GRB data \vs\ redshift compared to
a flat $\Lambda$CDM model with $\Omega_M = 0.369$,
and non-flat $\Lambda$CDM model with $\Omega_M = 0.416$
and $\Omega_{tot} = 1.115$, and an exponentially evolving
model in a $\Omega_M = 1$ model (Wright 2002).
\label{fig:dDM-vs-z-SNeGRB-15Jan07}}
\end{figure}

Schaefer (2007) has given a sample of 69 GRBs with redshifts
and distance moduli.  These values have been binned
as well, but with bin widths $ < 0.1(1+z_{min}+z_{max})$.  The
binned values are listed in Table \ref{tab:GRB}.  In
constructing the Table, a Milne model with a Hubble constant of
72 km/sec/Mpc was subtracted from the individual distance
moduli before minimizing the modified $\chi^2$.  
Figure \ref{fig:dDM-vs-z-SNeGRB-15Jan07} shows the binned
data from both the GRBs and the supernovae.

% \clearpage

\begin{figure}[t]
\plotone{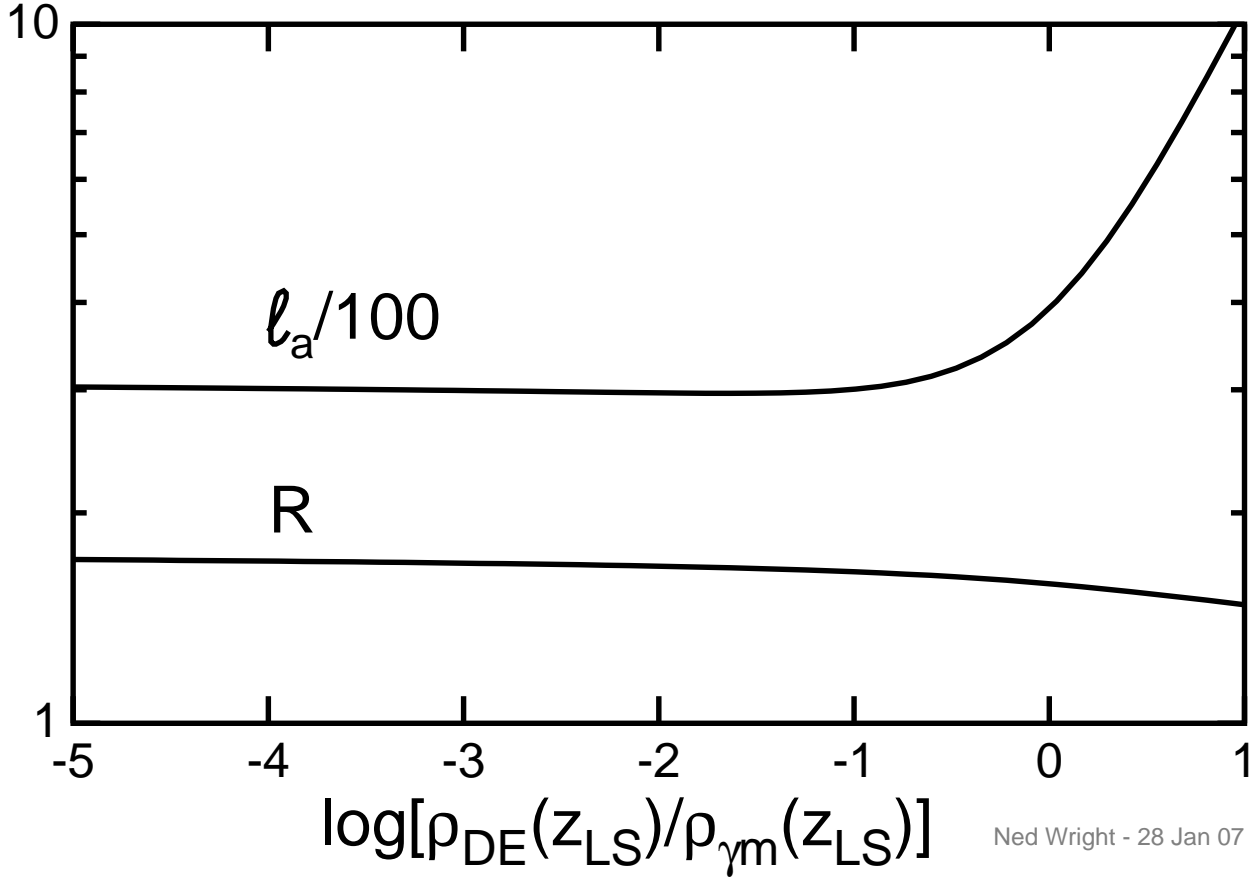}
\caption{The stretch parameter $R$ and the acoustic scale $\ell_a$
as a function of the ratio of dark energy density to matter plus
radiation density at $z_{LS}$.  These curves were evaluated
for flat models by varying $w^\prime$, setting $w_0 = -0.922 - 0.309 w^\prime$
and then adjusting $\Omega_M$ to minimize the $\chi^2$
from the supernovae fit plus the simple $\Omega_M = 0.296 \pm 0.029$ prior.
\label{fig:R-ell_a}}
\end{figure}

\begin{figure}[tp]
\plotone{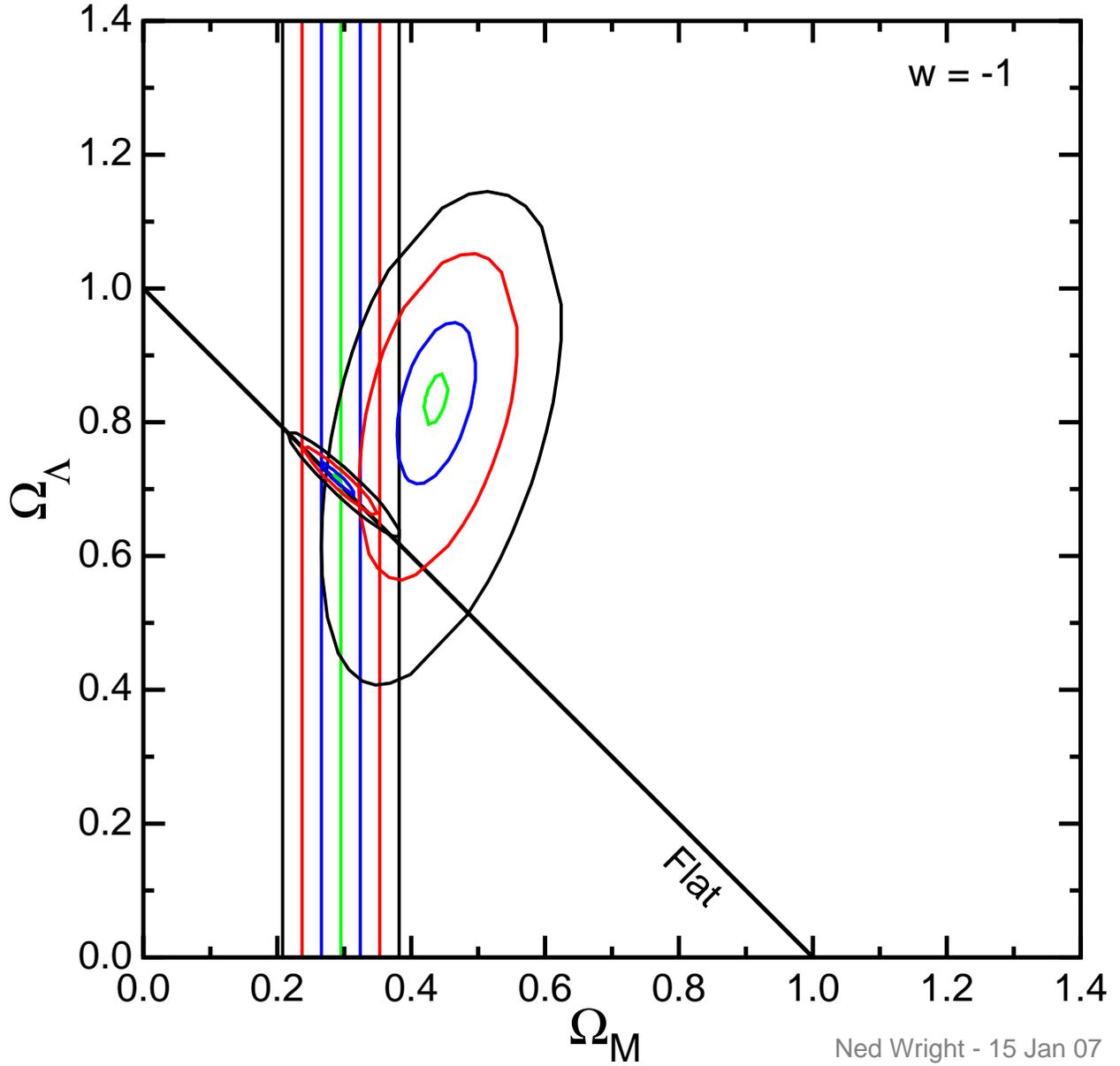}
\caption{Contours of $\chi^2$ in the $\Omega_M$-$\Omega_\Lambda$ plane
when the dark energy is assumed to be a cosmological constant with $w =
-1$.  The large ellipses use only the Hubble diagram data from the
supernovae and the GRBs.  The small ellipses near the ``Flat'' line use
only the CMB acoustic scale and the baryon acoustic distance ratio.
Contours are drawn for $\Delta\chi^2 = 0.1$, 1, 4, and 9.  The vertical
lines show -3 \ldots +3 $\sigma$ for the $\Omega_M$ prior derived from
$\Gamma$, $H_\circ$ and $\omega_M$.
\label{fig:Wm-Wv-both-CC-15Jan07}}
\end{figure}

\begin{figure}[tp]
\plotone{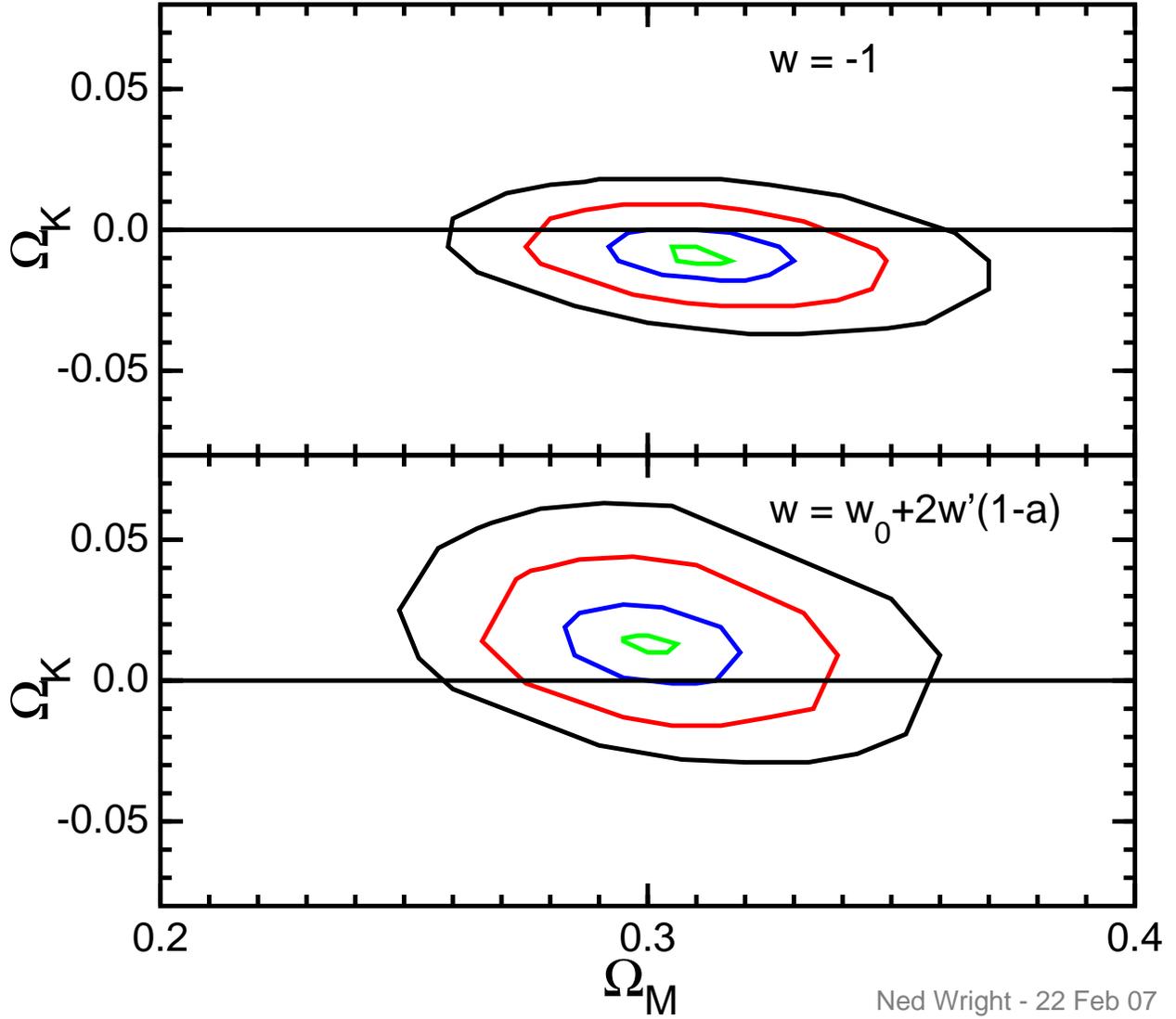}
\caption{Contours of $\chi^2$ in the $\Omega_M$-$\Omega_K$ plane.  In
the lower panel  the dark energy is allowed to vary as $w(z) =
w_\circ+2w^\prime(1-a)$, using all constraints and datasets.  A flat
model is consistent with the data, and the best fit is very slight
open.  In the upper panel $w = -1$, and a flat model is again
consistent but the best fit is slightly closed.
\label{fig:Wm-Wk-all-17Jan07}}
\end{figure}

\begin{figure}[tp]
\plotone{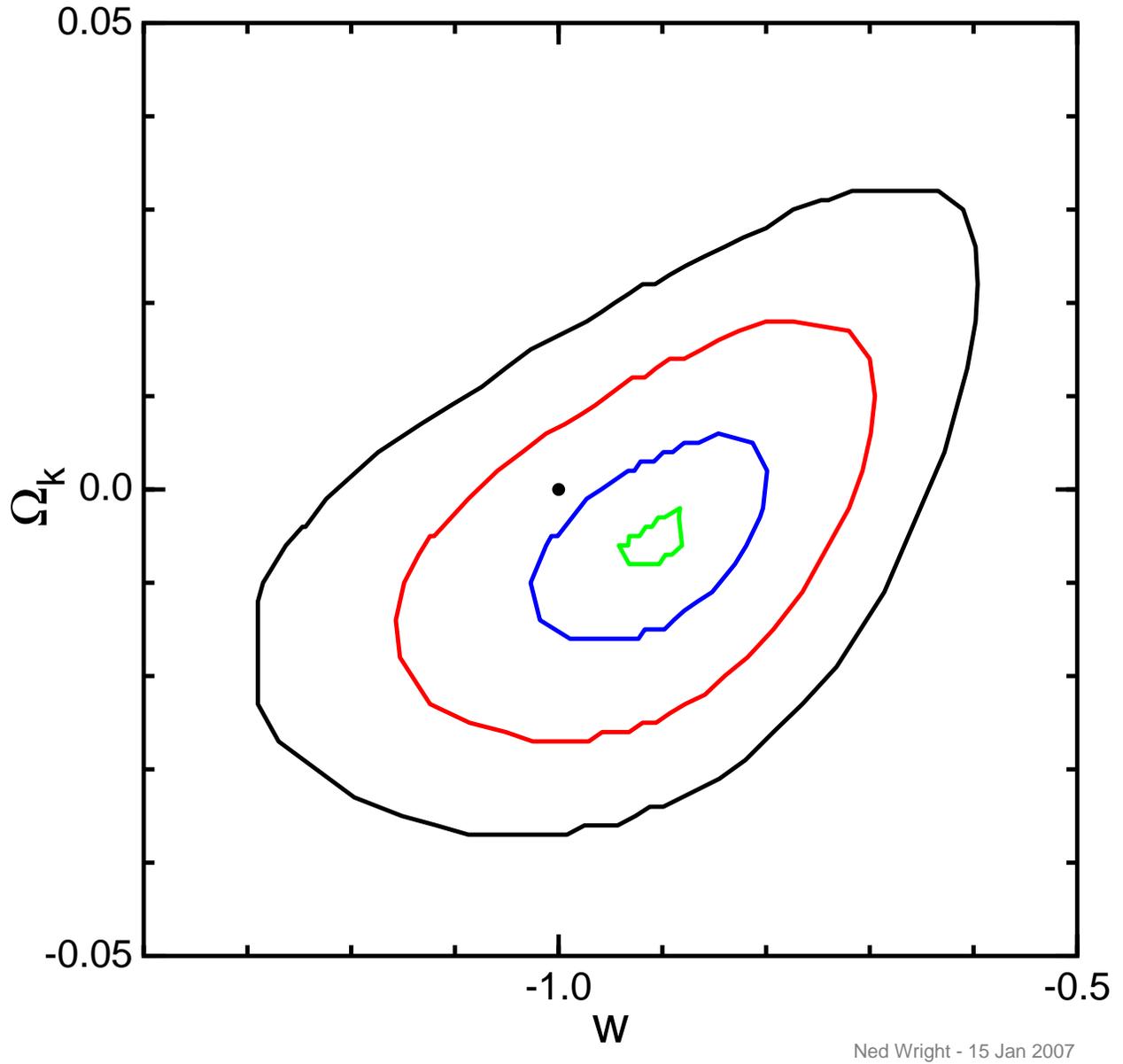}
\caption{Contours of $\Delta \chi^2$ in the $w$, $\Omega_K$ plane,
with $w^\prime$ fixed at zero and $\Omega_M$ adjusted to minimize
$\chi^2$ at each point.  The black dot shows the flat $\Lambda$CDM model,
which has 2 fewer free parameters and less than two units of $\Delta\chi^2$.
\label{fig:w-Wk-15Jan07}}
\end{figure}

\begin{figure}[t]
\plotone{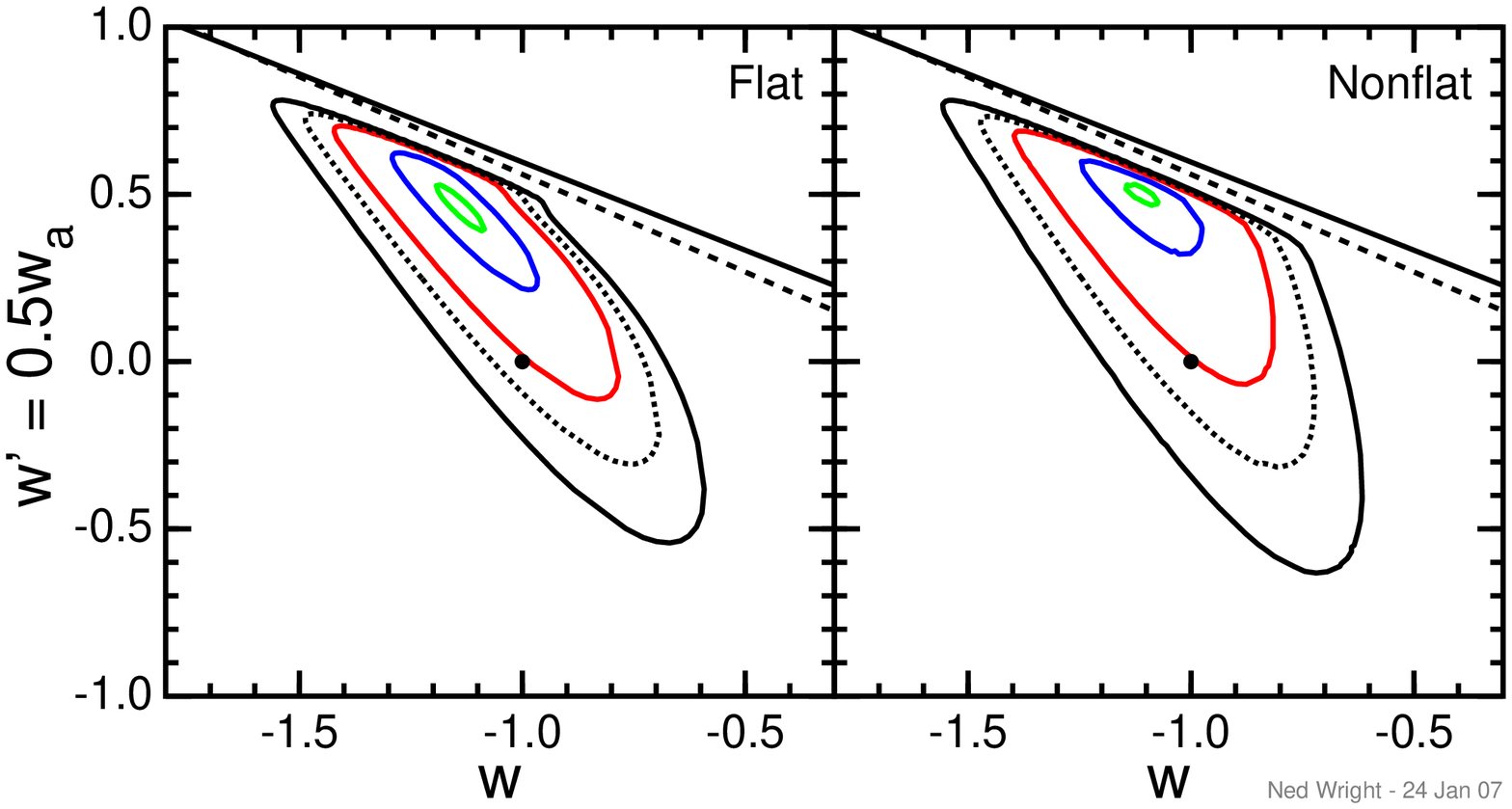}
\caption{Contours of $\Delta \chi^2$ in the $w$,  $w^\prime$ plane,
using all constraints and datasets.  
The dotted contour  shows the $\Delta\chi^2 = 6.17$ 
used in the DETF Figure of Merit.
On the left the curvature
$\Omega_K$ is fixed at zero and $\Omega_M$ is adjusted to minimize
$\chi^2$ at each point.  On the right both $\Omega_M$ and $\Omega_K$
are adjusted to minimize $\chi^2$ at each point.  The black dot shows
the cosmological constant $w = -1$.  The dashed diagonal line shows
where the dark energy is equal to the matter plus radiation density at
last scattering.  The solid diagonal line shows the 3 $\sigma$
limit on the stretch parameter $S^{BBNS}$.
\label{fig:w-wa-14Jan07}}
\end{figure}

% \clearpage

\subsection{Hubble Constant}

The Hubble constant largely cancels out in analyses of supernovae and
$\gamma$-ray burst distances.  But the Hubble constant does enter into
converting the $\Gamma$ parameter ($\Omega_M h$) into a prior on
$\Omega_M$.  The Hubble constant data used in this paper comes from
Freedman \etal\ (2001),  the DIRECT project double-lined eclipsing binary
in M33 (Bonanos \etal\ 2006), 
Cepheids in the nuclear maser ring galaxy M106 (Macri \etal\ 2006), and the
Sunyaev-Zeldovich effect (Bonamente \etal\ 2006).  These papers gave
values of $72 \pm 8$, $61 \pm 4$, $74 \pm 7$ and $77 \pm 10$ km/sec/Mpc.
Assuming that the uncertainties in these determinations are uncorrelated and
equal to 10 km/sec/Mpc after allowing for systematics, the
average value for $H_\circ$ is $71 \pm 5$ km/sec/Mpc.
This average is consistent with the $74 \pm 4(stat) \pm 5(sys)$~km/sec/Mpc
from Riess \etal\ (2005).

\subsection{Matter Density}

The matter density $\Omega_M h^2$ is fairly well determined by fitting
the CMB power spectrum.  In this paper the non-flat $\Lambda$CDM chain
(ocdm\_wmap\_1.txt) at the LAMBDA data center has been used to
determine average values for parameters.  This chain gives $\omega_M =
\Omega_M h^2 = 0.1289 \pm 0.0079$.  The baryonic density is also well
determined, with $\omega_B = \Omega_B h^2 = 0.02178 \pm 0.00072$.

\subsection{Large Scale Structure}

Large scale structure data comes from the ``big bend'' in the power
spectrum $P(k)$.  This has been measured by two large galaxy surveys:
the SDSS and the 2dF.  The SDSS gives a value for
$\Gamma = \Omega_M h = 0.213 \pm 0.0233$ (Tegmark \etal\ 2004),
while the 2dF gives (Cole \etal\ 2005)
\be
\Omega_M h = \Gamma_{true} = 0.168 \pm 0.016 + 0.3(1-n_s)
 + 1.2\Omega_\nu/\Omega_M
\ee
I assume that the Tegmark \etal\ value uses $n_s = 1$ and $\Omega_\nu =
0$, and that it has the same sensitivity to these parameters as the
2dF.  The combination of these two values gives a $\chi^2 = 2.6$ for 1
degree of freedom.  While this is higher than the expected value of 1,
it is certainly not high enough to trigger grave concerns.  The
neutrino density is uncertain but the minimal hierarchical mass pattern
gives $\Omega_\nu/\Omega_M \approx 0.004$ while the WMAP 3 year data
give $n_s = 0.951 \pm 0.017$ (Spergel \etal\ 2007).  With these values
the corrected weighted mean $\Gamma$ is $0.209 \pm 0.014$.

\subsection{Acoustic Oscillations}

The most important parameter from the CMB data is the acoustic scale.
In this paper the stretch parameter 
$R = \Omega_M^{1/2} H_\circ (1+z_{LS}) D_A(z_{LS})/c$ 
(Bond, Efstathiou \& Tegmark 1997;  Wang \& Mukherjee 2006) is not used, 
but the acoustic scale $\ell_a$ 
evaluated at the mean baryon and dark matter densities is used in its place.  
The acoustic scale is defined as (Page \etal\ 2003)
\be
\ell_a = \frac{\pi (1+z_{LS}) D_A(z_{LS})}{\int_0^{1/(1+z_{LS})} 
c_s da/(a\dot{a})}
\label{eq:ella}
\ee
where $z_{LS}$ is last scattering, $c_s$ is the sound speed, $a$ is the
scale factor, $\dot{a}$ is its time derivative, and $D_A$ is the
angular size distance.  The stretch parameter $R$ has a different
normalization and approximates the
denominator as $\propto \Omega_M^{-1/2}$, which is not a good
approximation when the dark energy is significant at $z_{LS}$.
$\ell_a$ is very well determined by the CMB data, with $\ell_a = 303.14
\pm 1.04$ in the non-flat $\Lambda$CDM chain.  
There are correlated deviations in $\omega_B$, $\omega_M$,
$\Omega_\Lambda$ and $\Omega_k$ that produce correlated
deviations in the numerator and denominator of Eq(\ref{eq:ella})
which cancel out in the ratio. 
In this paper the full
fit to the CMB data is not performed, but the baryon and CDM densities
are fixed at their mean values.  With this simplification, the
determination of the acoustic scale loosens to $\ell_a = 302.97 \pm
4.14$ because the deviations in the numerator are still present.  
The relative accuracy of $\ell_a$ with fixed $\omega_B$ and
$\omega_M$ is the same as the relative
accuracy of $R$.  The main reason to use $\ell_a$ instead of $R$ is
that it is clear how $\ell_a$ is modified when the dark energy is
important at last scattering.  Figure \ref{fig:R-ell_a} shows that $\ell_a$
shifts a great deal when the dark energy density is greater than the
radiation plus matter density at recombination, while the change
in $R$ is smaller and in the wrong direction.

The CMB also provides the matter density $\omega_M$ used
as part of the large scale structure analysis.   For models far from the
geometric degeneracy line, a full fit to the CMB power spectrum
will compensate for an incorrect distance to the surface of last
scattering by changing $\omega_M$ and $\omega_B$ to give 
a different value of the sound horizon at last scattering,
but this effect has not been been included here.  Ultimately
a full fit to all the data should be performed, and the CMB is
both the most informative and the most computationally intensive of
all the datasets.  Spergel \etal\ (2007) has some limits on $w$,
but not on the $w,w'$ plane, and does not include either the
Riess \etal\ (2007) and Wood-Vasey \etal\ (2007) increments
to the supernova data or the GRB data.

The second major input involving acoustic oscillations comes from the
baryon acoustic oscillations detected by Eisenstein \etal\  (2005).  In
this paper we use the ratio of the distance $D_V(0.35)$ at $z=0.35$ to
the tangential distance at last scattering,
\be
R_{0.35} = \frac{D_V(0.35)}{(1+z_{LS})D_A(z_{LS})}
\ee
This ratio is easily
computed even when the dark energy is significant at last scattering.
The ratio is also slightly more precise than the $A$ parameter since
the scatter induced by the uncertainty in $\omega_B$ and $\omega_M$
cancels out in the ratio.  The $A$ parameter also uses the approximation
that the sound travel distance is $\propto \Omega_M^{-1/2}$ which fails
when dark energy dominates early.  Finally, $A/R$ is exactly proportional
to $D_V(0.35)/[(1+z_{LS})D_A(z_{LS})]$.  Using both $A$ and the distance
ratio amounts to double counting the baryon acoustic oscillations.

The redshift of last scattering, $z_{LS}$, is essentially independent of
both the baryon density and the expansion rate because the electron
density is $n_e \propto H(z)/\alpha(T)$ while the optical depth is
$\tau \propto \sigma_T n_e c/H(z) \propto \sigma_T/\alpha(T)$ where
$\sigma_T$ is the scattering cross-section and $\alpha$ is the
recombination coefficient.  Hence $z_{LS} = 1089$ is used for all
calculations in this paper.  

\subsection{Big Bang Nucleosynthesis}

Steigman (2006) has analyzed the light element abundances and obtained
limits on a stretch factor $S^{BBNS} =
(1+\rho_{DE}/\rho_{\gamma+m+k})^{1/2}$.  The limits are $S^{BBNS} = 0.942 \pm
0.030$.  This means that at 3 $\sigma$ the dark energy density must be
less than 6.4\% of the radiation density at the redshift of
nucleosynthesis, $z \approx 10^9$.

\subsection{Expansion Rate at Last Scattering}

Zahn \& Zaldarriaga (2003) have shown
that even though $H(z)$ cancels out in determining $z_{LS}$, the
width in conformal time of the transition from opaque to transparent
can be determined using the CMB TT and TE power spectra, and
that this can give a stretch factor similar to the BBNS stretch factor.
Zahn \& Zaldarriaga interpret this in terms of varying Newton's
constant $G$, but it really depends on $\sqrt{G\rho}$.
Interpreting the limit in Zahn \& Zaldarriaga in terms
of the dark energy density instead of a changing $G$ gives 
$S^{LS} = (1+\rho_{DE}/\rho_{\gamma+m+k})^{1/2} = 1.749 \pm 0.471$ 
at $z_{LS}$.   With the first year WMAP data this improved to $1.04 \pm 0.24$ 
(Zahn, private communication).   

\section{ANALYSIS\label{sec:analysis}}

All of the analyses in this paper depend on the expansion history and
geometry of the Universe.  The expansion history can be calculated using
\be
\dot{a} = H_\circ \sqrt{\Omega_M/a + \Omega_R/a^2 + \Omega_K
+\Omega_{DE}[\rho_{DE}(z)/\rho_{DE}(0)]a^2}
\ee
The dark energy density as a function of redshift is computed using the
$w = w_0 + w_a(1-a)$ formula (Chevallier \& Polarski, 2001).  
Following Linder (2003) I set $w^\prime = w_a/2$.  This then gives
\be
\frac{\rho_{DE}(z)}{\rho_{DE}(0)} = (1+z)^{3+3w_0+6w^\prime}
\exp\left(\frac{-6w^\prime z}{1+z}\right)
\ee
The calculation of angular size and luminosity distances then
follows Wright (2006).

Note that the binned distance moduli were not used for the analysis.
Thus the Hubble constants used when producing the binned data tables
are irrelevant in the analysis.  But both the supernovae and the GRBs
have an associated ``nuisance'' parameter, ${\cal M}_{SN}$ and ${\cal
M}_{GRB}$, which are adjusted to give the best modified $\chi^2$ at
every position in the $\{\Omega_M, \Omega_k, w, w^\prime\}$ parameter
space.  Thus 
\bea 
\chi^2(\Omega_M, \Omega_k, w, w^\prime)  & = & \min_{{\cal M}} 
\sum_i  f((\mu_i-\mu_c(z_i))/\sigma_i) \nonumber \\
 & \mbox{with} & \nonumber \\ 
\mu_c(z) & = &  {\cal M} + 
5 \log(D_L(z; \Omega_M, \Omega_k, w, w^\prime)) 
\label{eq:HD}
\eea
The form in Eq(\ref{eq:HD}) is used twice, once for the 
supernovae giving $\chi^2_{SN}$ and once for the GRBs
giving $\chi^2_{GRB}$.  Contours of the
sum of these two terms are shown in 
Figure \ref{fig:Wm-Wv-both-CC-15Jan07}.

The acoustic scale quantities $\ell_a$ and $R_{0.35}$ give
a contribution to the overall $\chi^2$ of 
\bea
\chi^2_a(\Omega_M, \Omega_k, w, w^\prime) 
& = & 
\hat{f}(\ell_a(\Omega_M, \Omega_k, w, w^\prime)-\ell_{a,obs})/\sigma(\ell_a))
\nonumber \\
& + & 
\hat{f}(R_{0.35}(\Omega_M, \Omega_k, w, w^\prime)-R_{0.35,obs})/\sigma(R_{0.35}),.
\eea
where $\hat{f}$ is the function from Eq(\ref{eq:modchi}) with the transition
from quadratic to linear behavior set at $3 \sigma$:
$\hat{f}(x) = x^2$ for $|x| < 3$, or  $6|x|-9$ otherwise.

Contours of $\chi^2_a$ are shown in Figure \ref{fig:Wm-Wv-both-CC-15Jan07}.

Normally the data on $\Gamma = \Omega_M h$, $H_\circ$ and $\Omega_M h^2$
can be combined to give a prior on $\Omega_M$.  There is an overdetermined 
set of equations with variables $\Omega_M$ and $h$:
\be
\left(\begin{array}{cc} 1 & 1 \\ 1 & 2 \\ 0 & 1 \end{array}\right)
\left(\begin{array}{c} \ln(\Omega_M) \\ \ln(h) \end{array}\right) =
\left(\begin{array}{c} \ln(0.209) \pm (0.014/0.209) \\
\ln(0.1289) \pm (0.0079/0.1289) \\
\ln(0.71) \pm (0.05/0.71) \end{array}\right)
\ee
The least squares solution of these equations gives $H_\circ = 67.3 \pm 3.7$
km/sec/Mpc and $\Omega_M = 0.296 \pm 0.029$.
Contours showing this prior are the vertical lines
in Figure \ref{fig:Wm-Wv-both-CC-15Jan07}.

But the relationship between $\Gamma = \Omega_M h$ and the big bend in
$P(k)$ is derived assuming that matter and radiation are the only
significant contributors to the density at $z_{eq}$, the redshift of
matter-radiation equality.  Prior to $z_{eq}$, the Universe is
expanding faster than the free-fall time for matter perturbations, so
growth is suppressed for fluctuations that are inside the horizon
earlier than $z_{eq}$.  To allow for the possibility that dark energy
contributes, $z_{eq}$ is found using
\be
\Omega_M(1+z_{eq})^3 = \Omega_R(1+z_{eq})^4 + \Omega_K(1+z_{eq})^2 +
\Omega_{DE} (1+z_{eq})^{3+3w_0+6w^\prime}
\exp\left(\frac{-6w^\prime z_{eq}}{1+z_{eq}}\right)
\ee
The horizon at $z_{eq}$ is found using
\be
D_{eq} = \int_0^{1/(1+z_{eq})} \frac{c da}{a\dot{a}}
\ee
Then the effective value of $\Gamma$ is
 $\Geff = (1602\;\mbox{km/sec})/(H_\circ D_{eq})$.  
By finding the effective $\Geff$ at two different values
for $H_\circ$, which leads to two different values for $\Omega_R$ since
$\Omega_R h^2$ is fixed by the measurement of $T_\circ$,
the standard  $\Gamma = \Omega_M h$ can be replaced by a
modified power law function of $h$ which reduces to the standard form
except when the dark energy is significant near $z_{eq}$.  
For example,  in the track of models shown in Figure \ref{fig:R-ell_a},
when $\rho_{DE}/\rho_{\gamma m} = 0.138$ at $z_{LS}$ one has
$\Omega_M = 0.2983$, $w = -1.098$ and $w^\prime = 0.570$.
These values give $w(z) = 0.042$ at $z >> 10^3$, so the dark energy
density grows faster than the matter density but slower than the
radiation density at high $z$.
For these parameters one finds that 
$\Geff = 0.171704$ for $h = 0.675373$
and $\Geff = 0.188939$ for $h = 0.746402$.
Fitting a power law function of $h$ gives
$\Geff \approx 0.2499 h^{0.95657}$ instead of
the standard $0.2983 h$.  
For $h = 0.71$ this is a 16\% difference which
is quite significant compared to the 7\% precision of the
mean of the $\Gamma$'s from the SDSS and the 2dF.
When finding
$\chi^2$, a weighted mean estimate for $h$ is found using the
$\Gamma$, $H_\circ$ and $\omega_m$ priors.  This weighted mean
minimizes the $\chi^2$ contribution from these three priors, and this
minimum is added to the $\chi^2$ from the Hubble diagram, CMB, and the
BAO.  Therefore $h$ becomes a third nuisance parameter.  Then
\bea
\chi^2_\Gamma(\Omega_M,\Omega_k,w,w^\prime)  & = &
\min_h [\hat{f}((\Geff(\Omega_M,\Omega_k,w,w^\prime,h)
-\Gamma_{obs})/\sigma(\Gamma)) \nonumber \\
& + & \hat{f}(\Omega_M h^2 - \omega_{M,obs})/\sigma(\omega_M)
+ \hat{f}((h - h_{obs})/\sigma(h))]
\label{eq:Gamma}
\eea
is added to the overall $\chi^2$.
As the dark energy becomes significant at recombination, $\Omega_M$ has
to increase to keep the effective $\Gamma$ close to the observed value.
If we take the same $w = -1.098$ and $w^\prime = 0.570$ example
discussed above, and adjust $\Omega_M$ in flat models to minimize
$\chi^2_\Gamma$ alone, 
$\Omega_M$ increases by $1.5\sigma$ from 0.296 to 0.341.  
Thus a simple $\Omega_M$ prior is not correct when dark energy
is significant at $z_{LS}$.
Note that $\omega_{M,obs}$,
which is derived from the peak heights and trough depths in the CMB
angular power spectrum, will be affected by perturbations in the
dark energy density, but this effect has not been included in this
paper. 

It is simple to find the BBNS $S^{BBNS}$ parameter for any point in the 
$\{\Omega_M, \Omega_k, w, w^\prime\}$ parameter space, and
add an appropriate term to $\chi^2$.  In doing this calculation
I have used $h = \sqrt{\omega_M(CMB)/\Omega_M}$  for simplicity
instead of combining this analysis with the calculation of
$\chi^2_\Gamma$.
Since the desired value
for $S^{BBNS}$ is slightly less than one, while dark energy can 
only increase $S^{BBNS}$ above one,
the BBNS data act only as an upper limit on $w^\prime$.
The Zahn \& Zaldarriaga limit on dark energy at recombination is handled the
same way.    This gives a stretch factor term of
\bea
\chi^2_S & = & \hat{f}(S^{BBNS}(\Omega_M, \Omega_k, w, w^\prime)-S^{BBNS}_{obs})
/\sigma(S^{BBNS})) \nonumber \\
& + & 
\hat{f}(S^{LS}(\Omega_M, \Omega_k, w, w^\prime)-S^{LS}_{obs})/\sigma(S^{LS})).
\eea

The final form for the overall $\chi^2$ is
\be
\chi^2(\Omega_M, \Omega_k, w, w^\prime) = \chi^2_{SN} + \chi^2_{GRB}
+ \chi^2_a + \chi^2_\Gamma + \chi^2_S.
\ee
Note that a minimization over ${\cal M}_{SN}$ has already been done in
computing $ \chi^2_{SN}$ (see Eqn \ref{eq:HD}), 
a minimization over ${\cal M}_{GRB}$ has already
been done in computing $ \chi^2_{GRB}$ , 
and a minimization over $h$
has already been done in computing $ \chi^2_\Gamma$ (see Eqn \ref{eq:Gamma}).
Thus the overall $\chi^2$ is
\be
\chi^2(\Omega_M, \Omega_k, w, w^\prime) = \min_{h, {\cal M}_{SN},  {\cal M}_{GRB}}
\chi^2(\Omega_M, \Omega_k, w, w^\prime, h, {\cal M}_{SN},  {\cal M}_{GRB})
\label{eq:overall}
\ee
All of the contour plots in this paper except for 
Figure \ref{fig:Wm-Wv-both-CC-15Jan07} show contours of this combined
function.  Unplotted variables are either fixed at assumed values or
minimized over.  It is correct to remove nuisance parameters by
minimizing $\chi^2$ or maximizing the likelihood over them (Cash 1976).  

Marginalization by integrating the likelihood over the nuisance
parameters is wrong (Wright 1994), although it is allowable to marginalize by 
integrating over the {\it a posteriori} probability density function.  But this 
requires a correct prior distribution for the nuisance parameters.
The Monte Carlo Markov chain (MCMC) technique is a useful tool for performing
such integrals, but the MCMC thus requires correct priors.
As an example, consider Figure 20 of Spergel \etal\ (2007) which shows the
CMB constraint in the $(\Omega_M,\Omega_\Lambda)$ plane.  This should
have a uniform prior in  $(\Omega_M,\Omega_\Lambda)$, but the chain
was computed with a uniform prior in $(\Omega_\Lambda, \Theta_s)$, so the chain
weights had to be multiplied by the Jacobian of the transformation between
$(\Omega_\Lambda, \Theta_s)$ and $(\Omega_M,\Omega_\Lambda)$ before a
subset of the chain was selected for plotting.

Marginalization by minimization avoids the need for a prior on the
unplotted variables.  The plots in this paper show contours of the
likelihood, which are the same as contours of the {\it a posteriori}
probability density function if one assumes a uniform prior in the
plotted variables.

\begin{table}[t]
\caption{$\chi^2$ for fitting to 358 SNe, 69 GRBs, $\Gamma$, $H_\circ$,
$\omega_M$, and BBNS.  The parameters in the fit
are $w$, $w^\prime$, $\Omega_M$, $\Omega_K$, ${\cal M}_{SN}$,
${\cal M}_{GRB}$, and $h$.\label{tab:chi2}}
\begin{center}
\begin{tabular} {lccccc}
\tableline
Model type & $w_0$ & $w^\prime$ & $\Omega_M$ & $\Omega_K$ & $\chi^2$ \\
\tableline
flat $\Lambda$CDM     &  -1     &  0      &   0.306   &    0    &  427.905 \\
nonflat $\Lambda$CDM  &  -1     &  0      &   0.315   & -0.011  &  426.983 \\
nonflat constant $w$    &  -0.894 &  0      &   0.309   & -0.003  &  426.249 \\
flat varying $w$        &  -1.126 &  0.451  &   0.305   &    0    &  423.580 \\
nonflat varying $w$     &  -1.098 &  0.506  &   0.299   &  +0.015  &  422.856\\
\tableline
\end{tabular}
\end{center}
\end{table}

\section{DISCUSSION\label{sec:discussion}}

There are a large number of different cuts through the 4 dimensional
parameter space that can be plotted, and when different subsets of
the data are considered the number of plots multiplies rapidly.
The best fit $\chi^2$ for some of these combinations are listed
in Table \ref{tab:chi2}.
Figure \ref{fig:Wm-Wv-both-CC-15Jan07} shows the $\Omega_M$
\vs\ $\Omega_\Lambda$ plane when the dark energy is constrained
to be a cosmological constant with $w = -1$.  Three data subsets are
shown: the supernova and GRB Hubble diagrams,  the CMB plus
BAO acoustic scale data, and the $\Gamma$, $H_\circ$ and $\omega_M$
data.  Clearly the acoustic scale data give a strong
confirmation of the need for dark energy, and a much better
constraint on the curvature of the Universe.
Using all the data together gives the plot shown in Figure
\ref{fig:Wm-Wk-all-17Jan07}.  The best fit model is slightly closed
with $\Omega_{tot} = 1.011$ and $\Omega_M = 0.315$.
The best fit flat $\Lambda$CDM model has less than one more unit
of $\chi^2$ than the best fit non-flat $\Lambda$CDM model so there
is no evidence for spatial curvature from these fits.  Figure
\ref{fig:Wm-Wk-all-17Jan07} also shows the effect of allowing
$w(z)$ to vary.

Another way to see this is to plot $\chi^2$ \vs\ $w$ and $\Omega_K$,
as seen in Figure \ref{fig:w-Wk-15Jan07}.  In this plot $w^\prime$ is forced to 
zero, and $\Omega_M$ is adjusted to minimize $\chi^2$ at each point.
This plot is very similar to a comparable plot in Spergel \etal\ (2007).

The flat $\Lambda$CDM model
has only 5 more units of $\chi^2$ than a non-flat variable $w$
model with 3 more free parameters.  The probability of this occurring by
chance alone is over 16\%, so this improvement is not significant.

Plots of the $w$ \vs\ $w^\prime$ plane are shown in Figure
\ref{fig:w-wa-14Jan07},
with and without the assumption of a flat Universe.  It is obvious that
the contours would have extended to much higher values of $w^\prime$
if the nucleosynthesis and acoustic scale constraints had not been used.
The tilts of the ellipses below these cutoffs indicate that the pivot 
redshifts for the priors and
datasets used here are $z = 0.4$ and $z = 0.22$.  Values of
$w$ at the pivot redshift are close to $-0.9$ which is the best fit
when $w^\prime$ is forced to zero, as shown in Figure \ref{fig:w-Wk-15Jan07}. 

The Dark Energy Task Force (Albrecht \etal\ 2006) has defined a
Figure of Merit as one over the area of the $\Delta\chi^2 = 6.17$
contour in the $w,w_a$ plane.  Albrecht \etal\ said that this contour
is a 95\% confidence contour, which not exactly correct.  It is
actually a 95.4\% confidence contour to match the 2$\sigma$ confidence
of a one dimensional Gaussian (Albrecht, private communication).
The FoM for the flat case is 1.59 while for the non-flat case it
is 1.39, but both of these cases have run up against the constraints
against dark energy domination before last scattering.  These
constraints have given the non-flat case a bigger boost than the
flat case.  In fact the area of the $\Delta\chi^2 = 1$ contour is
smaller for the non-flat case than the flat case, which shows both
the inability of the current data to constrain $w,w_a$ and the
importance of using the constraints coming from high redshift
physics.

\section{CONCLUSION\label{sec:conclusion}}

All of the data considered here are consistent with a flat $\Lambda$CDM
model with a constant equation of state $w = -1$.  The current data,
even with well over 300 supernovae, are not adequate for measuring
a time variable equation of state with reasonable precision.  Serra,
Heavens \& Melchiorri (2007) and Davis \etal\ (2007) agree with
this conclusion, as do Liddle \etal\ (2006), Alam \etal\ (2006) and
Li \etal\ (2006) using earlier and smaller datasets.  The current
acoustic scale data, seen in the CMB and the baryon oscillations,
is giving more precise information about the expansion history of
the Universe, but without the dense redshift coverage provided by
the supernovae.  There appears to be a systematic deviation of the
supernovae data from the models around redshifts near 0.5, whose
origin is unknown.  Since the choice of data subsets affects the
size of this deviation it is probably an artifact.  Nesseris \&
Perivolaropoulos (2006) have found systematic differences between
the different data sources that went into Riess \etal\ (2007)
dataset, so artifacts are not unlikely.  Furthermore the scatter
between the distance determinations of identical supernovae by
different groups is unexpectedly large.  The GRB Hubble diagram is
not very precise but it appears to be consistent with the supernova
data.   The GRB Hubble diagram does help break the degeneracy between
$w \neq -1$ and $\Omega_K \neq 0$.  The 2dF and SDSS values for
$\Gamma$ differ by a slightly disturbing amount, so a new and
improved measurement of $\Gamma$ would be useful as a tie breaker.
The Hubble constant appears to be determined to better than 10\%
but independent new data would be quite valuable in pinning down
$\Omega_M$ in combination with $\Gamma$.  Better CMB data from
Planck should reduce the uncertainty in $\omega_B$ and $\omega_M$,
which will reduce the uncertainty in $R$ or $\ell_a$.  It is clear
that better data of many types will be needed to pin down $w(z)$.

\acknowledgements

We acknowledge the use of the Legacy Archive for Microwave Background Data 
Analysis (LAMBDA). Support for LAMBDA is provided by the NASA Office of 
Space Science.

\thebibliography

\bibitem[Alam \etal\ (2006)]{alam2006} Alam, U., Sahni, V., \& 
Starobinsky, A.~A.\ 2006, ArXiv Astrophysics e-prints, 
arXiv:astro-ph/0612381 

\bibitem[Albrecht \etal\ (2006)]{albrecht2006}
Albrecht, A. \etal\  2006, ArXiv Astrophysics e-prints, arXiv:astro-ph/0609591

\bibitem[Astier \etal\ (2006)]{astier2006}
 Astier, P. \etal\  2006, \aap, 447, 31 

\bibitem[Barger \etal\ (2006)]{barger2006} Barger, V., Gao, Y., \& 
Marfatia, D.\ 2006, ArXiv Astrophysics e-prints, arXiv:astro-ph/0611775 

\bibitem[Bonamente \etal\ (2006)]{bonamente2006} Bonamente, M., Joy, 
M.~K., LaRoque, S.~J., Carlstrom, J.~E., Reese, E.~D., \& Dawson, K.~S.\ 
2006, \apj, 647, 25

\bibitem[Bonanos \etal\ (2006)]{bonanos2006}
Bonanos, A.~Z. \etal\ 2006, \apj, 652, 313

\bibitem[Bond \etal\ (1997)]{bond/efstathiou/tegmark1997} Bond, J.~R., 
Efstathiou, G., \& Tegmark, M.\ 1997, \mnras, 291, L33 

\bibitem[Cash (1976)]{cash1976} Cash, W.\ 1976, \aap, 52, 307

\bibitem[Chevallier \& Polarski (2001)]{chevallier/polarksi2001}
Chevallier, M. \& Polarski, D. 2001, Int. J. Mod. Phys., D10, 213.

\bibitem[Cole \etal\ (2005)]{cole2005} Cole, S., \etal\ 2005, 
\mnras, 362, 505

\bibitem[Davis \etal\ (2007)]{davis2006} Davis, T. \etal\ 2007,  
ArXiv Astrophysics e-prints, arXiv:astro-ph/0701510

\bibitem[Eisenstein \etal\ (2005)]{eisenstein2005} Eisenstein, D.~J., 
\etal\ 2005, \apj, 633, 560 

\bibitem[Freedman \etal\ (2001)]{freedman2001} Freedman, W.~L.
\etal\ 2001, \apj, 553, 47 

\bibitem[Li \etal\ (2006)]{li2006} Li, H., Su, M., Fan, Z., 
Dai, Z., \& Zhang, X.\ 2006, ArXiv Astrophysics e-prints, 
arXiv:astro-ph/0612060

\bibitem[Liddle \etal\ (2006)]{liddle2006} Liddle, A.~R., 
Mukherjee, P., Parkinson, D., \& Wang, Y.\ 2006, \prd, 74, 123506

\bibitem[Linder(2003)]{2003PhRvL..90i1301L} Linder, E.~V.\ 2003, Physical 
Review Letters, 90, 091301

\bibitem[Macri \etal\ (2006)]{macri2006} Macri, L.~M., Stanek, 
K.~Z., Bersier, D., Greenhill, L.~J., \& Reid, M.~J.\ 2006, \apj, 652, 1133

\bibitem[Nesseris \& Perivolaropoulos(2006)]{nesseris2006} Nesseris, 
S., \& Perivolaropoulos, L.\ 2006, ArXiv Astrophysics e-prints, 
arXiv:astro-ph/0612653 

\bibitem[Page \etal\ (2003)]{page2003} Page, L., \etal\ 2003, 
\apjs, 148, 233

\bibitem[Riess \etal\ (2005)]{riess2005} 
Riess, A.~G. \etal\ 2005, \apj, 627, 579 

\bibitem[Riess \etal\ (2007)]{riess2007} Riess, A.~G., \etal\
2007, ArXiv Astrophysics e-prints, arXiv:astro-ph/0611572 

\bibitem[Schaefer (2007)]{schaefer2007} Schaefer, B.~E.\ 2006, ArXiv 
Astrophysics e-prints, arXiv:astro-ph/0612285 

\bibitem[Serra \etal\ (2007)]{serra2007} Serra, P., Heavens, A. \&
Melchiorri, A. 2007, ArXiv Astrophysics e-prints, arXiv:astro-ph/0701338

\bibitem[Spergel \etal\ (2007)]{spergel2006} Spergel, D.~N. \etal\
2007, ArXiv Astrophysics e-prints, arXiv:astro-ph/0603449

\bibitem[Steigman (2006)]{steigman2006} Steigman, G.\ 2006, ArXiv 
High Energy Physics - Phenomenology e-prints, arXiv:hep-ph/0611209

\bibitem[Tegmark \etal\ (2004)]{tegmark2004} Tegmark, M., \etal\ 
2004, \apj, 606, 702 

\bibitem[Wang \& Mukherjee(2006)]{wang/mukherjee2006} Wang, Y., \& 
Mukherjee, P.\ 2006, \apj, 650, 1

\bibitem[Wood-Vasey \etal\ (2007)]{woodvasey2006} Wood-Vasey, M. \etal\ 2007,  
ArXiv Astrophysics e-prints, arXiv:astro-ph/0701041

\bibitem[Wright (1994)]{wright1994} Wright, E.~L.\ 1994, CMB 
Anisotropies Two Years after COBE: Observations, Theory and the Future , 21

\bibitem[Wright (2002)]{wright2002} Wright, E.~L.\ 2002, 
ArXiv Astrophysics e-prints, arXiv:astro-ph/0201196 

\bibitem[Wright (2006)]{wright2006} Wright, E.~L.\ 2006, \pasp, 118, 1711

\bibitem[Zahn \& Zaldarriaga (2003)]{zahn/zaldarriaga2003} Zahn, O. \& 
Zaldarriaga, M. 2003, \prd, 67, 063002

\end{document}